\documentstyle[12pt,epsf]{article}
\begin{document}

\title{New type of beam size effect and the $W$-boson production
at $\mu ^+ \mu ^- $ collideres}

\author{K.~Melnikov\\{\em Institut fur
Physik,Universit\"{a}t Mainz}\thanks{ D 55099 Germany, Mainz,
Johannes Gutenberg Universit\"{a}t, Institut f\"{u}r Physik,
THEP, Staudinger weg 7; e-mail: melnikov@dipmza.physik.uni-mainz.de}
\\and\\V.G.~Serbo\\{\em Institut fur Theoretische Physik, Universit\"at
Leipzig }\thanks{Permanent address:
Novosibirsk State University, 630090, Novosibirsk, Russia; 
e-mail: serbo@phys.nsu.nsk.su
}}

\maketitle

\begin{abstract}
The cross section for the reactions of the type  
$\mu ^- \mu ^+ \to e \bar \nu _e X $
can not be calculated by the standard methods due to the $t$--channel
singularity in the physical region. In this letter we show that
accounting for the finite sizes of the colliding beams  
results in the regularization of this  singularity. 
The finite  cross section, which is obtained in this way, 
turns out to be linear proportional to the
transverse sizes of the colliding beams.
As an application of the above result, we calculate the cross section 
of 
the $W$ boson production at  $\mu^+\mu^-$ colliders in the reaction
$\mu ^- \mu ^+ \to e \bar \nu _e W^+ $.
\end{abstract}

\vspace{1.5cm}
\begin{center}
MZ-TH-96-02
\end{center}
\vspace{1cm}

\newpage

{\it 1. Introduction.}---
It is known since early 60's, that some high-energy 
processes can have a $t$-channel singularity in the physical 
region \cite {Peirles}. There is no commonly accepted solution for this 
problem. Typically, this situation occurs when initial particles in a given 
reaction are unstable and the masses of the final particles 
are such that the real decay of the initial particles can take place.

Recently, in the paper \cite {Ginz} it was stressed that this problem turned
out to be a practical issue for the 
reaction $\mu ^- \mu ^+ \to e \bar \nu _e  W^+$ . It is shown in 
\cite {Ginz} 
that the standard calculations lead to the infinite 
cross section in this case.
Indeed, if the invariant mass of the final $e\bar \nu _e$ system is smaller 
than the mass of the muon, the square of the momentum transfer in the 
$t$-channel $q^2$ can be 
both positive and negative, depending on the scattering angle. This results 
in the power-like singularity in the cross section 
$$
 d \sigma \propto \frac {dq^2}{|q^2|^2}
$$
and the standard 
calculations turn out to be impossible. 
It is therefore necessary to 
regularize this divergence in order to produce
definite prediction for the measurable number of events.

In this paper we show, {\it that accounting for the finite
sizes of the colliding beams 
 gives a finite answer for the processes with the $t$-channel 
singularity in the physical region}. This is the main result of our paper.
As an example, we consider reaction 
$\mu ^- \mu ^+ \to e \bar \nu _e W^+ $ and show that the actual 
cross section of 
this process is approximately $1$ fb for the typical transverse beam 
sizes of the order of $10^{-3}$ cm.

The beam size effect (BSE) at the high-energy colliders 
is well studied both experimentally and theoretically (for the review 
see Ref. \cite {Obzor}). For the first time 
this effect was observed at the VEPP-4 collider 
(Novosibirsk) in 1980-81 during the study of a single bremsstrahlung 
in the electron-positron collisions \cite {IYAF}. 
This year the BSE was observed at HERA in the reaction 
$ep \to ep\gamma $ \cite {HERA}.  In both cases the number of observed
photons was smaller than it was expected according to the standard 
calculations. The decreased number of photons is explained by the fact 
that impact parameters, which give  essential contribution to the 
standard cross sections of these reactions, are larger by 2-3 orders of 
magnitude compared to the transverse beam sizes.

In all the previous cases, when these effects were studied, the
results depended logarithmically on the beam sizes. Below we show, 
that in our case (the $t$-channel singularity in the physical region ) 
the cross section for a scattering process is 
{\it linear} proportional to the transverse 
sizes of the colliding beams.

{\it 2. Cross section of the reaction 
$\mu ^- \mu ^+ \to e \bar \nu _e X. $}---
We introduce the following notations: $s=(p_1+p_2)^2=4E^2$ 
is the square of the total 
energy in the center of mass frame, $m$ is the muon mass, 
$p_1^2 =p_2^2 =m^2$, $p_3$ is the 4-momentum of the final $e\bar \nu _e$ system,
$y=p_3^2/m^2 $ is the square of the 
invariant mass of the
$e\bar \nu _e$ system in units of the square of the muon mass, 
$q=p_1-p_3=(\omega, {\bf q})$ is the momentum transfer 
in the 
$t$--channel and  $x=\omega/E$ is the fraction of the initial muon energy 
transferred to the $t$-channel.

From simple kinematics it follows that
$$
q^2 = -{{{\bf q }} _\bot {}^2\over 1-x} + t_0,~~
-s(1-x)< q^2 < t_0, ~~~t_0= m^2 \, { x(1-x-y)\over 1-x}.
$$
Here  ${{\bf q }} _\bot$ is the component of the momentum $q$ 
which is transverse to the momenta of the initial muons. 
Note, that  $ t_0 >0 $ as far as $y < 1-x$ and that $q^2 =0$ at 
\begin{equation}
|{\bf q } _\bot| = q_\bot ^0 = m \sqrt{x(1-x-y)}. \label {Qtnul}
\end{equation}

Let us consider the region of $|q^2| < \Lambda ^2$ where $\Lambda \ll m$. 
In this region the main contribution comes from the diagram with the
exchange of the muonic neutrino in the $t$-channel (Fig.1).
Since for such $q^2$ the exchanged neutrino is almost real, the 
corresponding matrix element can be considerably simplified.

We present the matrix element $M$ in the form
\begin{equation}
M =-M_\mu~\frac {1}{q^2+i\epsilon}~M_{\nu \mu}.
\label {Matel}
\end{equation} 
Here $M_{\mu}$ is the matrix element for the muon decay and $M_{\nu \mu}$
is the matrix element for the $\nu _{\mu} \mu ^+ \to X$ process. 
In both of these subprocesses we take $q^2$ equal to zero.

\begin{figure}[htb]
\epsfxsize=6cm
\centerline{\epsffile{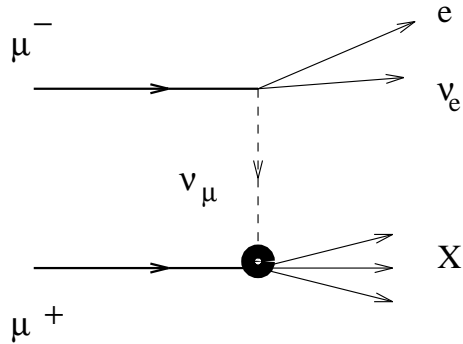}}
\caption[]{The Feynman graph for the reaction 
$\mu ^- \mu ^+ \to e \bar \nu _e X $, which gives the leading contribution 
in the region of small $|q^2|$.}
\end{figure}

Using the matrix elements for these subprocesses we express 
the cross section of the reaction 
$\mu ^- \mu ^+ \to e \bar \nu _e X $ through
the muon 
decay width $\Gamma$  
and the cross section $\sigma _{\nu \mu }$ 
of the $\nu _{\mu} \mu ^+ \to X$ process:

\begin{equation}
d\sigma = \frac {1}{\pi} x~m~d\Gamma~\frac {dq^2}{|q^2|^2}  
d\sigma _{\nu \mu},~~
 d\Gamma =2\Gamma (1-y)(1+2y)dxdy.
\label {sech}
\end{equation}
Let us call the coefficient in front of $ d\sigma _{\nu \mu} $ as
the number of neutrinos $dN_{\nu}$
$$
dN_{\nu}=\frac {1}{\pi} x~m~d\Gamma ~\frac {dq^2}{|q^2|^2}. 
$$

From the Eq.(\ref {sech}) it is clear that the standard 
calculation of this cross section turns out to be impossible due
to the power-like singularity 
since the point $q^2=0$ is within the physical
region for $y < 1-x$.
 
The main result of our study  of the BSE in the above process can 
be formulated as follows:

 {\it accounting for the BSE results in the following treatment of the
 divergent integral in the Eq. (\ref {sech})}:
\begin {equation}  
  B=\int \frac {dq^2}{|q^2|^2} \to \pi \frac {a}{q^0 _\bot}. 
\label {zajjavka}
\end {equation}

The exact expression for the quantity $a$ will be 
given below (see Eqs.(\ref {Bfact1}) and (\ref {Agauss})).
We just mention here that it is proportional to the 
transverse sizes 
of the colliding beams. For the identical round Gaussian 
beams with the mean square radii
$
\sigma _{ix}=\sigma _{iy}=\sigma _{\bot},~~i=1,2
$
this quantity is equal 
to 
$$ 
a=\sqrt{\pi} \sigma _\bot.
$$

Using this result in the expression for the number of neutrinos and 
integrating it over $y$, we arrive to the following spectrum of
the neutrino:
\begin{equation}
\frac {dN_\nu (x)}{dx} = \frac {\pi a}{2c\tau} f(x),~~
f(x)= \frac {24}{5\pi}
\sqrt{x(1-x)} \left(1+{22\over 9} x - {16\over 9} x^2 \right),~~
\int \limits_{0}^{1} f(x)dx =1.
\label {Eqne}
\end{equation}
Here $\tau$ is the life time of the muon at rest, $c\tau =660$ m.

After the number of  neutrinos (Eq.(\ref {Eqne})) 
is obtained,  the cross section 
for the reaction $\mu^-\mu^+ \to e\bar \nu _{e} X$ is given by the 
equation:
$$
d\sigma = dN _{\nu}(x)~d\sigma _{\nu \mu}(xs).
$$
Subsequent integration over $x$ can be performed without
further difficulties.

{\it 3. Derivation of the basic formula.}---
Now let us prove  our result  presented 
in the previous section (see Eq.(\ref {zajjavka})).

To begin with, we note that the standard notion of the cross
section is an approximation itself. As is well known,
it corresponds to the plane waves 
approximation for the initial and final particles. In the real experiments the 
particles are confined to the beams of a relatively small size and it 
is the collision of such beams that leads to the measurable number
of events. 

In order to arrive to a more general formulas for the description of the
scattering processes, we have to be able to describe the collisions of the 
wave packets instead of the plane waves. In view of the fact
that the movement of the particles inside a beam is 
quasiclassical, a simple and efficient technique
for taking into account the beam size effects in the actual calculations
has been developed \cite {Obzor}.

Below we present some results from the Ref. \cite {Obzor}, which are 
essential for our discussion. For simplicity, we neglect the energy and 
angular spread of the particles in the colliding beams.

Let us remind that in the standard approach the number of events $N$
is the product of the cross section $\sigma $ and the luminosity $L$:
\begin{equation}
dN=d\sigma~L,~~~~ d\sigma =
\frac {(2\pi)^4\delta(p_1+p_2-P_f)|M|^2}
{4p_1p_2} \prod _{f} \frac {d^3p_f}{(2\pi)^32E_f},
\label {Standard}
\end{equation}
$$ 
L=2\int n_1({\bf r},t)~n_2({\bf r},t)d^3rdt
$$ 
where $2=|{\bf v}_1-{\bf v}_2|$ for the head-on collision of the 
ultrarelativistic beams. 
The quantities $n_i({\bf r},t)$ are the particle densities of the beams.

The transformation from the plane waves to the colliding wave packets
results in the following changes. The squared matrix element $|M|^2$
with the 
initial state in the form of the plane waves with the momenta ${\bf p_1}$
and ${\bf p_2}$ transforms to the product: 
\begin{equation}
|M|^2 \to M_{fi}M_{fi'}^{*}. \label {Trans}
\end{equation}
Here the initial state $|i \rangle $ is the direct product of 
the plane waves with
the momenta ${\bf p}_1+\frac {1}{2}{\mbox {\boldmath $\kappa $ }} $
and ${\bf p}_2- \frac {1}{2}{\mbox {\boldmath $\kappa $ }}$, while the
initial state $|i' \rangle $ is the direct product of 
the plane waves with
the momenta ${\bf p}_1- \frac {1}{2}{\mbox {\boldmath $\kappa $ }}$
and ${\bf p}_2+\frac {1}{2}{\mbox {\boldmath $\kappa $ }}$.
Instead of the luminosity $L$ the number of events starts to depend
on the quantity 
$$
L({\mbox {\boldmath $\varrho $ }} )=
2\int n_1({\bf r},t)~n_2({\bf r}+{\mbox {\boldmath $\varrho $ }},t)d^3rdt
$$
through the following formula:
$$  
dN=d\sigma ({\mbox {\boldmath $\kappa $ }} ) 
L({\mbox {\boldmath $\varrho $ }} ) \exp (i{\mbox {\boldmath $\kappa $ }} 
 {\mbox {\boldmath $\varrho $ }})~
\frac {d^3\kappa d^3\varrho}{(2\pi)^3},
$$
\begin{equation}
d\sigma({\mbox {\boldmath $\kappa $ }} )=
\frac {(2\pi)^4\delta(p_1+p_2-P_f)~M_{fi}M_{fi'}^{*}}{4p_1p_2} 
\prod _{f} \frac {d^3p_f}{(2\pi)^32E_f}.
\label {Compl}
\end{equation}

The characteristic values of $ \kappa $ are of the order of the inverse
beam sizes, i.e.  $\kappa \sim 1/\sigma _\bot$. Usually this quantity
is much smaller than the typical scale for the variation of the
matrix element with respect to the initial momenta. In this case we can
put $\kappa =0$ in $d\sigma({\mbox {\boldmath $\kappa $ }} )$ which
results in the standard expression for the number of events 
Eq.(\ref {Standard}).
Otherwise, one should 
use complete formulas which take into account the effect of the 
finite beam sizes.

In view of the discussion given in the previous section, 
this is indeed the situation which occurs in our case. Now we want to show 
how the finite result for the number of events can be obtained starting 
from the complete formula Eq.(\ref {Compl}).

Let us 
first define the `` observable cross section'' by the relation
$d\sigma =dN/ L$
where $L$ is the standard luminosity.
By writing the number of events in such a way, we push the BSE to the 
quantity $d\sigma $.

The study of the matrix element of the discussed process (\ref {Matel})
suggests that the only quantity sensitive to the small variation of the 
initial momenta is the denominator of the neutrino propagator.
Henceforth, the transformation (\ref {Trans}) reduces to:
$$
\frac {1}{|q^2|^2} \to \frac {1}{t}~\frac {1}{t'}.
$$
Here
$$
t=q^2-{\mbox {\boldmath $\kappa $ }} {\bf q } _\bot+i\epsilon,~~
t'=q^2+{\mbox {\boldmath $\kappa $ }} {\bf q } _\bot-i\epsilon .
$$
In the expression for $t$ and $t'$ we only keep the terms which are 
linear in ${\mbox {\boldmath $\kappa $ }}$.

As a result the quantity $B$ (cf. Eq. (\ref {zajjavka})) transforms
to:
\begin{equation}
B=\int \frac {dq^2}{t~t'}~\frac {L({\mbox {\boldmath $\varrho $ }})}{L}~
\exp (i{\mbox {\boldmath $\kappa $ }} {\mbox {\boldmath $\varrho $ }})
\frac {d^3\kappa d^3\rho}{(2\pi)^3}. \label {Bfact}
\end{equation}
To proceed further,  we extend the 
region of integration over $q^2$ up to $\pm \infty$ and take
${\bf q } _\bot$ in the point where $q^2=0$ \footnote { More detailed 
discussion of this calculation will be given elsewhere \cite {Future}.}. 
After that integrations become simple and we obtain:

\begin{equation}
B = \frac {\pi}{q_\bot ^0}a,~~ a=\int \limits _{0}^{\infty} d\varrho \frac 
{L(\varrho {\bf n})}{L}, ~~~ {\bf n} = \frac {{\bf q } _\bot ^0}{q_\bot ^0}.
\label {Bfact1}
\end{equation}
This completes the proof of the Eq.(\ref {zajjavka}).

At high-energy colliders the distribution of particles in colliding 
beams can be often considered as Gaussian.
In this case $L(\varrho {\bf n})$ equals:
\begin{equation}
L(\varrho {\bf n}) = L\exp \Big \{ -\varrho ^2 \Big ( 
\frac {\cos^2\varphi}{2a_x^2}+\frac {\sin^2\varphi}{2a_y^2} \Big ) \Big \},~~
{\bf n}=(\cos\varphi,\sin\varphi)
\end{equation}
where $a_x^2=\sigma _{1x}^2+\sigma _{2x}^2$ and
$a_y^2=\sigma _{1y}^2+\sigma _{2y}^2$. 

This results in the following expression for $a$:
\begin{equation}
a= \sqrt{\frac {\pi}{2}}~\frac { a_x~a_y}{\sqrt{a_y^2 \cos^2\varphi+
 a_x ^2 \sin^2\varphi }}.
\label {Agauss}
\end{equation}

{\it 4. The cross section of the process
$\mu^-\mu ^+ \to  e \bar \nu _e W^+$.}---
Now we consider a special example, taking $X=W^+$.
The cross section of the 
reaction $ \nu _{\mu} \mu ^+ \to W^+$ is equal to
\begin{equation}
\sigma(\nu _{\mu} \mu ^+ \to W^+)=12\pi^2\frac {\Gamma(W\to \mu \nu)}{M}
\delta (xs-M^2)
\end{equation}
where $\Gamma(W\to \mu \nu)=0.22$ GeV is the partial W decay width and
$M=80.2$ GeV is the $W$ boson mass.
The integration over the fraction of the neutrino energy $x$ becomes 
trivial and finally we obtain:
\begin{equation}
\sigma(\mu^-\mu ^+ \to  e \bar \nu _e W^+)=\sigma _0
\frac {\pi a}{2c\tau}~x_0f(x_0),~~~
x_0=\frac {M^2}{s},
\end{equation}
$$
\sigma_0=\frac {12\pi^2}{M^2}\frac {\Gamma(W\to \mu \nu)}{M}=19.7 
{\mbox{nb}}.
$$

For numerical estimates we take
$a=\sqrt{\pi}\sigma_\bot$ (which corresponds to the case of the round 
identical Gaussian beams) with
$\sigma _\bot=10^{-3}$ cm (see Ref.\cite {Palm}). 

\begin{figure}[htb]
\epsfxsize=8cm
\centerline{\epsffile{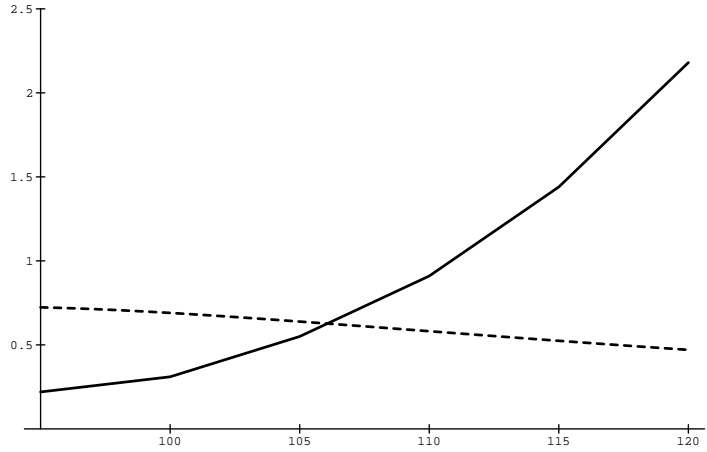}}
\caption[]{``Standard'' (solid line) and ``non-standard'' (dashed line)
contributions to the cross sections (fb) of the reactions 
$\mu ^- \mu ^+ \to e \bar \nu _e W^+ $  as a function of the
total energy $\sqrt {s}$ , GeV. The standard 
contribution 
is evaluated with the cut $-q^2 > m^2$.}
\end{figure}

This cross section reaches the maximum of $0.76$ fb for $\sqrt{s}=93$ GeV.
For larger energies this cross section decreases as $s^{-3/2}$.  

First, let us compare this ``non-standard'' contribution with the 
``standard'' one. By the ``standard'' contribution we mean the 
cross section of the same reaction calculated by the standard rules 
excluding the region of the final phase space
where $q^2~>~-m^2$. This contribution was calculated 
\cite {Private1}
using the CompHEP package \cite {Boos}.
The comparison of both contributions is shown in Fig.2.
It is seen that the non-standard piece dominates up to the energies
$\sqrt {s} \approx 105$ GeV.

Second, we compare our ``non-standard'' cross section with the 
cross section for the single $W$ boson production  in the reaction
$\mu^-\mu^+ \to \mu ^- \bar \nu _{\mu} W^+$. 
The latter is completely standard process since it has no $t$--channel
singularity.
The reasonable estimate 
of its cross section can be quickly obtained with the help of 
the equivalent photon approximation.  It gives the cross section
$\approx 1$ fb at $\sqrt {s} \approx 95 $ GeV (where it  almost
coincides with our non-standard cross section). At higher energies
the process $\mu^-\mu^+ \to \mu ^- \bar \nu _{\mu} W^+$ strongly dominates
as compared with the discussed process
$\mu^-\mu ^+ \to  e \bar \nu _e W^+$. 

{\it In summary}, we have shown that the $t$-channel singularity in the
physical region can be regularized by taking into account the finite sizes
of the colliding beams. Our results, being applied to the reaction
$\mu^-\mu ^+ \to  e \bar \nu _e W^+$, show that the non-standard 
contribution for this reaction dominates up to the energies $\sim 105$ GeV.

K.M. is grateful to  Graduiertenkolleg ``Teilchenphysik'', Universit\"at
Mainz for support and to E.~Sherman for a number of fruitful conversations.
V.G.S. acknowledges support of the S\"achsisches Staatsministerium f\"ur
Wissenschaft und Kunst and of the Russian Fund of Fundamental Research.
We thank I.F.~Ginzburg and G.L.~Kotkin for valuable discussions. 
We are grateful to E.E.~Boos and A.E.~Pukhov
for providing us with the results of the CompHEP calculations.

\end{document}